\documentstyle[12pt]{article}
\begin{document}
\setlength{\unitlength}{1mm}
{\hfill  JINR E2-94-118, March 1994}
\begin{center}
{\Large\bf On Exact Integrability of 2D Poincar\'e Gravity}
\end{center}
\begin{center}
{\large\bf Sergey Solodukhin$^{\ast}$}
\end{center}
\begin{center}
{\bf Bogoliubov Laboratory of Theoretical Physics,
Joint Institute for Nuclear Research,
Head Post Office, P.O.Box 79, Moscow, Russia}
\end{center}
\vspace*{2cm}
\begin{abstract}
We consider the 2D Poincar\'e gravity and show its exact integrability.
The choice of the gauge is discussed. The Euclidean solutions on
compact closed differential manifolds are studied.
\end{abstract}
\vspace*{2cm}
\vskip 4cm
\noindent $^{ \ast}$ e-mail: solodukhin@main1.jinr.dubna.su
\newpage

It is well known problem in two-dimensions how to determine the
dynamical description of gravity. One  possible way is to use
the gauge approach previously developed for
four-dimensional case [1]. This leads to description of gravity in terms of
zweibeins $e^a=e^a_\mu dx^\mu$ and the Lorentz connection one-form
$\omega^a_b=\omega^a_{\ b,\mu} dx^\mu$ as independent variables.
The theory  described by quadratic in curvature and torsion action
was suggested [2] and integrability of equations of motion was
investigated in the conformal gauge. The light cone gauge was studied
in [3]. The different aspects of quantization of the model was
considered in [4]. In [5] was shown that choosing appropriate
coordinate system on 2D manifold $M^2$ one finds the exact solution
of the gravitational equations. In this note we give more transparent
and strict proof of the exact integrability and describe some issues
which were shadowed in previous consideration [5].

\bigskip

 In two dimensions the gauge gravity is described in terms of zweibeins
$e^{a} = e^{a}_{\mu} dz^{\mu}, a=0,1 $. The 2D metric on the surface $M^{2}$
has the form $g_{\mu \nu}=e^{a}_{\mu} e^{b}_{\nu} \eta_{ab} $
$\eta_{ab}= diag (+1,(-1)^s)$
 and Lorentz
connection one-form $\omega^{a}_{\ b} = \omega \varepsilon^{a}_{\ b}, \  \omega
=
\omega_{\mu} dz^{\mu} \  (\varepsilon_{ab} =- \varepsilon_{ba}, \
\varepsilon_{01}=1)$.
For $s=0$ we have a Euclidean and for $s=1$ a Minkowskian signature.
The curvature and torsion two-forms
are: $R=d\omega, \ T^{a}=de^{a} + \varepsilon^{a}_{\ b}
\omega \wedge e^{b}$.

   The dynamics of gravitational variables $( e^{a}, \omega)$ is determined by
the action [2,5]:
\begin{equation}
S_{gr}= \int\limits_{M^{2}}^{} {\alpha \over 2} \ast T^{a} \wedge T^{a} +
{\beta \over 2} \ast R \wedge R + (-1)^s {\lambda^{} \over 4} \varepsilon_{ab}
e^{a} \wedge e^{b}
\end{equation}
where $\ast$ is the Hodge dualization and $\alpha,\beta,\lambda$ are arbitrary
constants.
The theory (1) with $\beta=\lambda=0$ was analyzed in [6] and was shown to
describe
some kind of topological gravity. Here we consider the case when these
constants are
not zero and fix $\beta=1$.
Generally one may add to (1)
the term proportional to $R$. But this term is boundary one
 and it does not affect the equations of motion.

   It is convenient to consider variables $\rho = \ast R$ and $q^{a} = \ast
T^{a}$.
Variation of action (1) with respect to zweibeins $e^{a}$ and Lorentz
connection $\omega$ leads to the following equations of motion [5]:
\begin{equation}
d \rho =- {\alpha} q^{a}\varepsilon_{ab}e^{b}
\end{equation}
\begin{equation}
\nabla q^{a} = {(-1)^s \over 2{\alpha} }
\Phi (q^2,\rho)
 \varepsilon^{a}_{\ b} e^{b},
\end{equation}
where $\nabla q^{a} \equiv dq^{a} + \omega \varepsilon^{a}_{\ b}q^{b}$;
here $q^{2}=q^{a}q^{b} \eta_{ab}$. In (2), (3) the following
notation was introduced:
$\Phi (q^2,\rho)=\rho^{2} +\alpha q^{2}- \alpha^{2} - \Lambda \alpha$, where
$\Lambda= {\lambda \over \alpha}- {\alpha}$.

It follows from eq.(3) that
\begin{equation}
dq^2= {(-1)^{s+1} \over \alpha^2} \Phi (q^2,\rho) d \rho
\end{equation}
It is differential equation for $q^2 (\rho)$, the solution is function [5]
$$
q^2(\rho)=- {1 \over \alpha} (\rho+(-1)^{s+1} \alpha)^2 +\Lambda +\epsilon
e^{(-1)^{s+1} {\rho \over \alpha}}
$$
where $\epsilon$ is the integrating constant.

It should be noted that below analysis is not dependent on the concrete
form of functions $\Phi (q^2, \rho)$ and $q^2 (\rho)$. In Euclidean case, of
course,
our analysis is valid not for any $\rho$ but only in the regions where
$q^2(\rho) \geq 0$.

\bigskip

The general solution of eqs.(2-3) is one of two types. The first one is
the space-time with zero torsion squared $q^2 \equiv 0$.
{}From eqs.(2)-(3) we obtain then that  torsion is zero $q^a \equiv 0$ and
curvature
is constant: $\rho=\pm \sqrt{\lambda}$. The second type of solutions is
characterized
by that the torsion $q^a$ is not zero identically on $M^2$. To analyze it
let us now consider the following one-form:
\begin{equation}
\xi={1 \over q^2} e^{(-1)^{s+1}{\rho \over \alpha}} q_a e^a
\end{equation}

It is direct consequence of gravitational equations (2)-(3) that
this one-form is closed $d\xi=0$. Indeed, straightforward calculation gives:
\begin{equation}
d\xi={ e^{(-1)^{s+1}{\rho \over \alpha}} \over q^2} ( \nabla q_a e^a +q_a
\nabla e^a+(-1)^{s+1}
({1 \over \alpha}- {\Phi \over \alpha^2 q^2})d\rho \wedge q_a e^a)
\end{equation}
{}From eq.(2) we obtain:
\begin{equation}
d\rho \wedge (q_a e^a)=\alpha q^2 V
\end{equation}
where $V={1 \over 2} e^a \varepsilon_{ab} e^b$ is volume 2-form and we used the
identity
$e^b \wedge e^c= (-1)^s \varepsilon^{bc}{1 \over 2} \varepsilon_{\alpha\beta}
e^{\alpha}\wedge e^\beta$. We also have that $\nabla e^a =(-1)^s q^a V$.

Thus substituting eqs.(3) and (7) into eq.(6) we obtain that
$$
d\xi=0.
$$

Now assuming that the 2D space-time $M^2$ has trivial topology (namely
the first cohomology group is trivial: $H^1(M^2)=0$) there exists globally
defined one-valued scalar field $\phi$ such that
\begin{equation}
\xi=d \phi
\end{equation}
If $H^1(M^2) \neq 0$,
say $M^2$ is cylinder, such a field $\phi$ still exists and is just angle
variable(which changes in the interval $0 \leq \phi \leq 2\pi$)
corresponding to the relevant nontrivial circle. Below we assume $M^2$ to be
topologically trivial or direct product $M^2=S^1 \oplus R^1$.

\bigskip

{}From eqs. (3), (5), (8) we obtain that
\begin{eqnarray}
&&q^a \varepsilon_{ab} e^b=-{ 1 \over \alpha} d\rho \nonumber \\
&&q_a e^a =q^2 e^{(-)^s{\rho \over \alpha}} d \phi
\end{eqnarray}
One can see that fields $(\rho,\phi)$ determine the natural coordinate
system on $M^2$.

Equations (9) are easily solved with respect to the orthonormal basis $e^a$:
\begin{equation}
e^a=q^a e^{(-1)^s{\rho \over \alpha} } d\phi +{(-1)^s \over \alpha q^2}
\varepsilon^a_{\ b}
q^b d\rho
\end{equation}
and metric $ds^2=e^a_\mu e^b_\nu \eta_{ab} dx^\mu dx^\nu$ takes the form
\begin{equation}
ds^2 =q^2 e^{(-1)^s{2\rho \over \alpha}} (d\phi)^2 +{(-1)^s \over \alpha^2 q^2}
(d\rho)^2,
\end{equation}
where function $q^2(\rho)$ is found from eq.(4).

\bigskip

The initial system (2-3) is differential equations for zweibeins
$e^a$ and the Lorentz connection $\omega$ as unknown functions.
Let us now find the Lorentz connection $\omega$. One can do this in two
different
(but equivalent) ways. The first one is to solve the equation

$$
de^a +\varepsilon^a_{\ b} \omega \wedge e^b =T^a
$$
with respect to $\omega$ assuming that zweibeins $e^a$ are already known and in
 coordinates $(\rho, \phi)$ expressed accordingly to (9).

The other way is to consider the equation
\begin{equation}
q^a \varepsilon_{ab}\nabla q^b=(-1)^{s+1}{\Phi \over 2\alpha} q_a e^a
\end{equation}
obtained  by multiplication eq.(3) on $q^b \varepsilon_{ba}$.

Let for definiteness $q^2 >0$ (for $s=1$), then one can introduce the new field
$\theta$ in following way
\begin{eqnarray}
&&q^0=q \cos{\theta} , \ q^1=q \sin{\theta} \ \ for \ \ s=0 \nonumber \\
&&q^0=q \cosh{\theta} , \ q^1=q \sinh{\theta} \ \ for \ \ s=1
\end{eqnarray}
where $q \equiv \sqrt{q^2}$.
Then one has
\begin{equation}
q^a \varepsilon_{ab} dq^b =q^2 d \theta
\end{equation}
Solving eq.(12) with respect to $\omega$ one finally obtains
\begin{equation}
 \omega + (-1)^{s+1}d \theta =-{\alpha \over 2} (q^2)'_{\rho} e^{(-1)^s {\rho
\over \alpha}}
d \phi
\end{equation}
This completes the proof of exact integrability of eqs.(2-3).

\bigskip

It should be noted that to this moment everything was Lorentz invariant.
Under local Lorentz rotations on angle $\eta$ the Lorentz connection
one-form $\omega$ and $\theta$ transform as follows
\begin{equation}
\omega \rightarrow \omega +(-1)^s d \eta , \ \theta \rightarrow \theta + \eta
\end{equation}
So the eq.(15) is indeed Lorentz invariant.

\bigskip

Thus we obtain that in coordinates $(\rho,\phi)$ solution of equations
(2-3) is given by (10) and (15) and depends on an arbitrary field
$\theta$. This arbitrariness is just reflection of  the underlying local
Lorentz symmetry.

Now we may use this freedom and choose the gauge fixing. One can suggest
the following choices.

\bigskip

{\it A. $\phi=F(\theta)$}.

$F$ is arbitrary analytic and monotonic function. Then $d\phi=B_0 (\theta)
d\theta$
and from eqs.(10)(15)
takes the form
\begin{eqnarray}
&&ds^2 =q^2 e^{(-1)^s{2\rho \over \alpha} } B^2_{0}(\theta) d \theta^2
+{(-1)^s \over \alpha^2 q^2} d \rho^2 \nonumber \\
&&\omega=-((-1)^{s+1}+{\alpha \over 2}(q^2)'_{\rho} e^{(-1)^s{\rho \over
\alpha}}
B_{0}(\theta)) d \theta
\end{eqnarray}
It is the gauge which was used in the [5] and expressions (17) are exactly that
of
obtained in [5].

\bigskip

{\it B. $\theta=0$}.

This gauge is equivalent to the condition $q^1=0$. Then
$q=q^0$ and
expression for $\omega$ is given by
\begin{equation}
\omega=-{\alpha \over 2} (q^2)'_\rho e^{(-1)^s{\rho \over \alpha}} d \phi
\end{equation}
It is worth noting that in this gauge the equations (2-3) are essentially
simplified and solution
is obtained at once. This gauge seems to be very useful in solving
eqs.(2-3) when the coupling to matter is taken into account.

\bigskip

In paper [7] was suggested the general 2D Poincar\'e gravity
\begin{equation}
S=\int_{M^2}^{} U(\rho, q^2) {1 \over 2} e^a \varepsilon_{ab} \wedge e^b
\end{equation}
with the Lagrangian $U$ being arbitrary function of curvature $\rho$ and
torsion $q^2$. Variation of (19) with respect to $e^a$ and $\omega$
gives the equations (both for $s=0$ and $s=1$):
\begin{eqnarray}
&&d(U'_{\rho})=-2 U'_{q^2} q^a \varepsilon_{ab} e^b \nonumber \\
&&\nabla (U'_{q^2} q^a) ={1 \over 2}(\rho U'_{\rho} +2q^2
U'_{q^2} -U) \varepsilon^a_{\ b} e^b
\end{eqnarray}
One can prove the exact integrability of these equations along the same line
as before. We will do this in quite different way than in the [7].

Indeed, let us introduce new variables $\bar{q}^a, \bar{\rho}$:
\begin{equation}
\bar{q}^a=V'_{q^2} q^a , \ \bar{\rho}={1 \over 2} U'_\rho
\end{equation}
We assume that transformation $(q^2,\rho) \rightarrow (\bar{q}^2, \bar{\rho})$
is correct and relevant Hessian is not zero. This means, in particular, that
there exists the inverse transformation such that one gets
$q^2=q^2(\bar{q}^2, \bar{\rho}), \rho=\rho(\bar{q}^2, \bar{\rho})$.
Then the eqs.(20) are rewritten in terms of the variables $\bar{q}, \bar{\rho}$
as follows
\begin{eqnarray}
&&d \bar{\rho} =- {\alpha} \bar{q}^{a}\varepsilon_{ab}e^{b} \nonumber \\
&&\nabla \bar{q}^{a} = -{1 \over 2 }
\Phi (\bar{q}^2,\bar{\rho})
 \varepsilon^{a}_{\ b} e^{b},
\end{eqnarray}
Function $\Phi (\bar{q}^2, \bar{\rho}) $ is  $(U-\rho U'_\rho-2q^2 U'_{q^2})$
under condition that $q^2,\rho$ are expressed in terms of $\bar{q}^2,
\bar{\rho}$.
One can see that eqs.(22) take the form (2-3). The above proof of integrability
of system (2-3) was made for general function $\Phi$. Really the form of the
solution (10), (15) is not dependent on the concrete form of the
function $\Phi$. The later influences only on the dependence
$\bar{q}^2(\bar{\rho})$
as solution of differential equation $d\bar{q}^2 / d\bar{\rho} =
\Phi (\bar{q}^2, \bar{\rho})$.
So
we have some type of universality [8] and
the exact solution of eqs.(22) again takes the form (10), (15)

Thus, the 2D Poincar\'e gravity gives us an unique example when the
equations are integrated and the exact general solution takes
analytically simple form.

\bigskip

The previous analysis concerned the non-compact manifolds. Let us consider
now the Euclidean compact closed two-dimensional differential manifold $M^2_g$
of genus $g$.
It is worth noting that one-form,
dual to $\xi$:
\begin{eqnarray}
&&\ast \xi ={1 \over q^2} e^{-\rho \over \alpha} q^a \varepsilon_{ab} e^b
\nonumber \\
&&=- {d\rho \over \alpha q^2(\rho)} e^{-{\rho \over \alpha}}
\end{eqnarray}
is obviously closed:

$$
d(\ast \xi)=0
$$

Thus, we obtain that $\xi$ and $\ast \xi$ are both closed one-forms.
For closed compact differential manifold it means that one-form $\xi$ (or/and
$\ast \xi$)
is the harmonic one and hence it represents of the first cohomology
group of $M^2_g$: $\xi \in H^1(M^2_g)$. If $M^2$ is topologically sphere
($g=0$), i.e.
$dim H^1 (M^2)=0$, then $\xi=0$. This means that torsion is identically zero on
$M^2$. Hence, only solution of the first type ($q^a \equiv 0$ and
$\rho=-\sqrt{\lambda}$)
is realized.

Generally, one should note that really the form $\ast \xi$ is exact:

$$
\ast \xi= d \psi
$$

where

$$
\psi=-\int_{}^{} e^{-{\rho \over \alpha}}{d\rho \over \alpha q^2(\rho)}
$$
Since curvature $\rho$ is globally defined one-valued on $M^2$ function,
$\psi$ is one-valued function too. Though, $\psi$ can take infinite values at
points where
$q^2(\rho)=0$. Hence $\xi=0$ for any genus $g$.

Thus, in general case the following Theorem is valid.

{\it Theorem}: Let $M^2_g$ is compact closed two-dimensional differential
manifold
of genus $g$. The equations (2), (3) being considered on $M^2_g$ can have only
solutions of the first type, i.e. the torsion is identically zero $q^a=0$ and
curvature is constant: $\rho=- \sqrt{\lambda}$ for $g=0$, $\rho=0$ for
$g=1$, $\rho= \sqrt{\lambda}$ for $g >1$.

The value of constant curvature is determined in agree with the Euler number of
$M^2_g$, $\chi=2(g-1)$.

However, this doesn't exhaust all possible solutions in the class of closed,
not necessary
differential, manifolds: the metric (11) can describe compact closed manifolds
with conic singularities at points where $q^2(\rho)=0$.
The complete analysis of the Euclidian solutions will be given elsewhere.

\bigskip

I would like to thank M.O.Katanaev for useful discussions inspired the
writing of these notes. I also thank Prof.F.W.Hehl and Dr.Yu.N.Obukhov
for kind hospitality at University of Cologne. This work was supported
in part by the grant 94-02-03665-a of Russian Fund of Fundamental
Investigations.

\end{document}